\documentclass{article}
\usepackage{multirow, amsmath, amsthm, graphicx, amssymb, amsfonts, latexsym, enumerate, amscd}
\usepackage{listings}
\RequirePackage[numbers]{natbib}
  
\begin{document}
	
	   \title{SKEWED PROBIT REGRESSION - IDENTIFIABILITY, CONTRACTION
		AND REFORMULATION} 
	\author{Janet van Niekerk
		and H{\aa}vard Rue}
\maketitle
	

   

        
        \begin{abstract}
            Skewed probit regression is but one example of a
            statistical model that generalizes a simpler model, like
            probit regression. All skew-symmetric distributions and
            link functions arise from symmetric distributions by
            incorporating a skewness parameter through some skewing
            mechanism. In this work we address some fundamental issues
            in skewed probit regression, and more genreally skew-symmetric distributions or skew-symmetric link
            functions. 
            
            We address the issue of identifiability of the skewed probit model
            parameters by reformulating the intercept from first
            principles. A new standardization of the skew link
            function is given to provide and anchored interpretation
            of the inference. Possible skewness parameters are
            investigated and the penalizing complexity priors of these
            are derived. This prior is invariant under
            reparameterization of the skewness parameter and
            quantifies the contraction of the skewed probit model to
            the probit model.
            
            The proposed results are available in the \textit{R-INLA}
            package and we illustrate the use and effects of this work
            using simulated data, and well-known datasets using the link as well as the likelihood.
        \end{abstract}
        

        
        \newpage     
        \section{INTRODUCTION}  
        \label{intro}
        Skew-symmetric distributions have acclaimed fame due to their
        ability to model skewed data, by introducing a skewness
        parameter to a symmetric distribution, through some skewing
        mechanism. In the preceding decades, an abundance of skewed
        distributions has been proposed from the basis of symmetric
        distributions, like the skew-normal
        \cite{ohagan1976,azzalini1985}, skew-t \cite{azzalini2003} and
        more generally skew-elliptical distributions
        \cite{genton2004}. In each of these skew distributions, an
        additional parameter is introduced that indicates the
        direction of skewness or alternatively, symmetry.
        
        With the introduction of the additional parameter, the
        inferential problem can become more challenging. The
        identifiability of the parameters and the existence of the
        maximum likelihood estimators (MLEs) are issues to keep in
        mind. In the Bayesian paradigm, the choice of a prior for the
        skewness parameter emerges. Either way, the inference of the
        skewness parameter is crucial in evaluating the
        appropriateness of the underlying (skewed) model.
        
        A continuous random variable $X$, follows a skew-normal (SN)
        distribution with location, scale and skewness(shape)
        parameters $\xi, \omega$ and $\alpha$, respectively, if the
        probability density function (pdf) is as follows:
        \begin{equation}
            g(x)=\frac{2}{\omega}\phi\left( \frac{x-\xi}{\omega}
            \right)
            \Phi\left[\alpha \left( \frac{x-\xi}{\omega} \right) \right],
        \label{skewnormpdf}
        \end{equation}
        where $\alpha\in \Re$, $\omega>0$, $\xi\in\Re$, and
        $\phi(\cdot)$ and $\Phi(\cdot)$ are the density and cumulative
        distribution function (CDF) of the standard Gaussian
        distribution, respectively. Denote by $G(x)$ the CDF of the
        skew-normal density.

        The parameterisation in \eqref{skewnormpdf} poses difficulties
        since the mean and variance depends on $\alpha$, as
        $E[X]=\xi+\omega\delta\sqrt{2/\pi}$ and
        $V[X]=\omega^2\left(1-2\delta^2/\pi\right)$, where
        $\delta=\alpha/\sqrt{1+\alpha^2}$. This implies that inference
        for $\alpha$ will also influence the inference for the mean
        and variance, since both are functions of $\alpha$.

        A similar challenge arises in the binary regression framework
        where the skew-normal link function is used as a
        generalization of probit regression, namely skewed probit
        regression. The need for asymmetric link functions have been
        noted by \cite{chen1999}. In binary regression, asymmetric
        link functions are essential in cases where the probability of
        a particular binary response approaches zero and one at
        different rates. In this case, a symmetric link function will
        result in substantially biased estimators with
        over(under)estimation of the mean probability of the binary
        response, due to the different rates of approaching zero and
        one (see \cite{czado1992} for more details on this issue).
        Skewed probit regression is an extension of probit regression,
        where covariates are transformed through the skew-normal CDF
        instead of the standard normal CDF.
        
        Here, it might not be intuitive when the skewed link function
        is more appropriate than the symmetric link function. The
        estimate of the skewness parameter could provide some insights
        into this, only if the inference of the skewness parameter is
        reliable and interpretable.
        
        Regarding the inference of the skewness parameter, $\alpha$ in
        \eqref{skewnormpdf}, being it in the skewed probit regression
        or the skew-normal distribution as the underlying response
        model (which are conceptually the same estimation setup),
        various works have been contributed, most of them dedicated to
        the skew-normal response model framework. The identifiability
        of the parameters in the skew-normal response model was
        investigated by \cite{genton2012} (and skew-elliptical in
        general), \cite{otiniano2015} (for finite mixtures) and
        \cite{castro2013} (for extensions of the skew-normal
        distributions). For binary regression, identifiability of the
        parameters was considered by \cite{lee2019} where some issues
        concerning identifiability were raised. We address the
        identifiability problem from a first principles viewpoint, so
        that the parameters are identifiable, even with weak
        covariates, hence adding to \cite{lee2019}.
        
        In the skew-normal response model, the bias of the MLEs is a
        well-known fact (see \cite{sartori2006} for more details). For
        small and moderate sample sizes, the MLE of the skewness
        parameter could be infinite with positive probability and the
        profile likelihood function has a singularity as the skewness
        parameter approaches zero, as noted early on by
        \cite{azzalini1985} (see also \cite{liseo1990}). Some
        approaches to alleviate this feature of the skew-normal
        likelihood function have been proposed, including
        reparameterization of the model by \cite{azzalini1985} using
        the mean and variance (instead of location and scale
        parameters), or using a Bayesian framework by \cite{liseo2006}
        (default priors) and \cite{bayes2007} (proper priors). Also,
        \cite{sartori2006} used the work of \cite{firth1993} to
        propose an adjusted (penalized) score function for frequentist
        estimation of the skewness parameter. A penalized MLE approach
        for all the parameters, including the skewness parameter, is
        presented by \cite{azzalini2013}. Bias-reduction regimes were proposed by \cite{maghami2020}.
        
        From a Bayesian viewpoint, various priors for the skewness
        parameter have been proposed such as the Jeffrey's prior
        \cite{liseo2006}, truncated Gaussian prior
        \cite{arellano2007}, Student t prior and approximate Jeffery's
        prior \cite{bayes2007}, uniform prior \cite{azevedo2011},
        probability matching prior \cite{cabras2012}, informative
        Gaussian and unified skew-normal priors \cite{canale2016} and
        the beta-total variation prior \cite{dette2018}. All of these
        Bayesian approaches, with the exception of the latter, are
        based on somewhat arbitrary prior choices for mainly
        mathematical or computational convenience. These priors (as
        many others) are not invariant under reparameterization of the
        skewness parameter. The beta-total variation prior presented
        by \cite{dette2018} is based on the total variation from the
        symmetric Gaussian model to the skew-normal model, viewing the
        skewness parameter as a measure of perturbation. This prior is
        indeed invariant under one-to-one transformation of the
        skewness parameter.
        
        Amongst the many works on the skew-normal response model, it
        seems that the genesis of the skew-normal model has been
        neglected. The skew-normal model was introduced by
        \cite{azzalini1985} as an (asymmetric) extension of the
        Gaussian model. The motivation for this extension is found in
        data. When data behaves like the Gaussian model, but the
        profile of the density is asymmetric, the skew-normal model
        might be appropriate. Conversely, we need an inferential
        framework wherein the skew-normal model would contract (or
        reduce) to the Gaussian model, in the absence of sufficient
        evidence of non-trivial skewness. The priors mentioned before
        do not provide a quantification framework with which the
        modeler can understand, and subsequently control this
        contraction. To achieve this, we need to consider the model
        (either skewed probit regression or the skew-normal response
        model) from an information theoretic perspective. Then we can
        construct a prior with which the quantification of contraction
        (or not) can be done, and used as a translation of prior
        information from the modeler to the model.
        
        In this paper we address some issues (identifiability,
        standardizing, skewness parameters) prevalent in skewed-probit
        regression in Section \ref{skewedsection} and construct the
        penalized complexity (PC) prior for the skewness parameter of
        the link function (which is translatable to the skew-normal
        response model) in Section \ref{pcsection}. This PC prior is
        implemented in the \textit{R-INLA} \cite{rue2009} package for general
        use by others. We use a numerical study to illustrate the
        solutions proposed in Section \ref{skewedsection} and apply
        the PC prior to simulated and real data in Sections
        \ref{simstudy} and Section \ref{realsection}. The paper is
        concluded by a discussion in Section \ref{disc} in which we
        sketch the wider applicability of this work and contributions
        to the wider skew-symmetric family.
        
        \section{SKEWED PROBIT REGRESSION AND ISSUES}\label{skewedsection}

        We consider skewed probit regression as an extension of probit
        regression, where the link function is the skew-normal CDF
        instead of the standard normal CDF. We formulate skewed probit
        regression that can include random effects like spline
        functions of the covariates, spatial and/or temporal effects.
        For this paper, we assume the following structure. From a
        sample of size $n$, the responses $\pmb{y}_{n\times 1}$ are
        counts of successful trials out of $N_{n\times 1}$ trials and
        hence we assume a Binomial distribution with success
        probability $p$. We gather all $m$ covariates in
        $\pmb{X}_{n\times m}$ and use these to build an additive
        linear predictor, defined as $\pmb\eta_{n\times 1}$. So then,
        \begin{eqnarray}
          y_i&\sim & \text{Binomial}(N_i,p_i)\notag\\
          p_i&=&G(\eta_i), \quad i=1,\ldots,n\label{skewprobitgen}
        \end{eqnarray}
        where $G(\cdot)$ is the CDF of the Skew-Normal  that
        depends on $(\xi, \omega, \alpha)$. The linear predictor
        $\eta_i $ is an additive linear predictor defined as follows,
        \begin{equation}
            \eta_i = \beta_0 + \pmb{\beta}'\pmb{X}_i +
            \sum_{k=1}^Kf^k(\pmb{Z}_i),\label{linpred}
        \end{equation}
        where $\pmb X$ and $\pmb Z$ are the covariates for the fixed
        and random effects, respectively, the functions $\{f^k(.)\}$
        are random effects like spatial, spline, temporal effects with
        hyperparameters $\pmb\theta$.

        \subsection{Issue 1 - Standardizing the link function}

        With the aim of standardizing the link function,
        \cite{lee2019} assumed $\xi=0,\omega=1$, similar to
        \cite{bazan2006} and many others. Initially, the idea behind
        this choice feels intuitive since the skew probit link is an
        extension of the probit link through the skewness parameter.
        However, the $(0,1)$ parameter values of the probit link
        should not be naively copied to the skewed probit link. The
        choice, $\xi=0,\omega=1$ implicitly concedes that a
        skew-normal density \eqref{skewnormpdf} with mean
        \begin{equation*}
            E[X]=\alpha\sqrt{\frac{2}{\pi(1+\alpha^2)}},
        \end{equation*} and variance 
        \begin{equation*}
            V[X]=1-\frac{2\alpha^2}{\pi(1+\alpha^2)},
        \end{equation*}
        is used to calculate the probability of success, for all
        $\alpha$. This essentially implies that for different skewness
        parameter values, different means and variances are used. This
        way of standardizing is a parameter-based method, instead of
        the intended property-based method like in the probit link. We
        do not expect the assumption $\xi=0, \omega=1$ to work well
        since the mean and variance are not anchored and can attain
        many values based on different values of $\alpha$.
        
        We posit that the mean and the variance (properties of the
        link) should be fixed, like in the probit case, instead of the
        skew-normal location and scale parameters. This is analogous
        to the idea of the centered parametrization of the skew-normal
        density and mentioned by \cite{bazan2010}.

        We propose the link function $F(y|\alpha)$ that is the CDF of
        the Skew-Normal density \eqref{skewnormpdf} scaled to have
        zero mean and unit variance for all values of $\alpha$. That
        is,
        \begin{displaymath}
            F(y|\alpha) =\int_{-\infty}^{y} f(x|\alpha)\,dx
        \end{displaymath}
        where
        \begin{equation}
            f(x|\alpha)=\frac{2}{\omega(\alpha)}\phi\left(
              \frac{x-\xi(\alpha)}{\omega(\alpha)} \right)
            \Phi\left[\alpha \left(
                \frac{x-\xi(\alpha)}{\omega(\alpha)} \right) \right], 
            \label{skewnormpdf2}
        \end{equation}
        \begin{equation*}
            \xi(\alpha) = -\omega\alpha\sqrt{\frac{2}{\pi(1+\alpha^2)}},
        \end{equation*}
        and 
        \begin{equation*}
            \omega(\alpha) = \sqrt{\left(1-\frac{2\alpha^2}{
                      \pi(1+\alpha^2)}\right) ^{-1}}.
        \end{equation*}
        This provides an anchored link function with zero mean and
        unit variance, for all $\alpha$. If this standardization is not
        used then an arbitrary unknown scale is introduced to the
        model, with no means of recovering it. By fixing the mean and
        variance, we have a better understanding of the properties of
        the link and we do approach the probit case in the
        neighborhood of $\alpha=0$.
        
        \subsection{Issue 2 - The quantile intercept and
            identifiability of parameters}\label{secint}

        The identifiability of the parameters in skewed probit
        regression were first investigated by \cite{lee2019}. They
        showed that without the presence of a continuous covariate,
        the intercept $\beta_0$, and skewness parameters are not
        identifiable. This is expected due to the traditional
        definition of the skewed probit model \eqref{skewprobitgen}
        and \eqref{linpred}. We rectify the formulation of the skewed
        probit regression intercept, by introducing the quantile
        intercept, and subsequently solve this issue of
        non-identifiability by returning to first principles.
        
        In simple linear regression, the intercept is used to
        calculate the expected value of the linear predictor without
        any effect from covariates. In probit regression, the
        intercept contains information about the probability of the
        event, without the effects from covariates. The value of the
        intercept should not provide any information about the other
        parameters in the model.
        
        However, when we introduce a skewness parameter to a symmetric
        family to formulate a skew-symmetric link then we are
        fundamentally changing the meaning of what is traditionally
        called the intercept of the linear predictor, i.e. $\beta_0$
        in \eqref{linpred}.

        Consider probit regression with one centered covariate $X$,
        \begin{equation*}
            p = \text{Prob}[Y=1] = \Phi(\beta_0 + \beta_1X).
        \end{equation*}
        Now if $\beta_1X = 0$, then 
        \begin{equation*}
            q = \text{Prob}[Y=1] = \Phi(\beta_0),
        \end{equation*}
        which implies that $\beta_0$ is the $q^{\text{th}}$ quantile
        of the standard Gaussian distribution. There is thus a
        one-to-one relationship between $q$ and $\beta_0$. When
        $\beta_1\neq 0$, then $\text{Prob}[Y=1]$ changes because of
        $\beta_1X$, without affecting $\beta_0$, because $\Phi$
        remains the same function. In this sense, $\beta_0$ is
        uninformative for $\beta_1$.

        Conversely, consider skewed-probit regression from
        \eqref{skewprobitgen} and \eqref{skewnormpdf2},
        \begin{equation*}
            p=\text{Prob}[Y=1] = F(\beta_0 + \beta_1X|\alpha).
        \end{equation*}
        Here, $\beta_0$ should, in the same way, be uninformative for
        $\beta_1$. This does not hold because the dependence of
        $\alpha$. We can ensure this, if
        \begin{equation*}
            q=\text{Prob}[Y=1] = F(\beta_0|\alpha)
        \end{equation*}
        is constant for varying $\alpha$, which is the case if
        $\beta_0$ is defined as the $q^{\text{th}}$ quantile of the
        distribution with CDF $F$. Therefore, we reformulate $\beta_0$
        as
        \begin{equation}
            \beta_0(q,\alpha) = F^{-1}(q|\alpha),
            \label{quantint}
        \end{equation} 
        so $\beta_0$ is the $q^{\text{th}}$ quantile of $F(.|\alpha)$.
        The quantile level $q$ is now the unknown intercept-parameter
        instead of $\beta_0$.

        Note that there is (generally) not a one-to-one relationship
        between $\beta_0$ and $q$ since the $q^{\text{th}}$ quantile
        depends on $\alpha$. In this new formulation, the intercept as
        defined implicitly by $q$, provides no information about
        $\beta_1$ and parameters of $F(\eta_i|\alpha)$ are
        identifiable. We return in \ref{subsecconfounding} to a
        numerical study of this issue.

        This formulation might seem surprising at first sight, but in the case of a symmetric link,
        the intercept is the quantile of a distribution with fixed
        (no) skewness. In the case of the probit or identity links for
        example, this formulation will reduce to the usual intercept
        parameter since in these cases there is a one-to-one
        relationship between $\beta_0$ and $q$.
        
        In terms of implementation in \textit{R-INLA}, the new formulation of
        the skew normal model in terms of $q$ is available and
        subsequently, the prior distribution for $q$ can be derived
        from a corresponding informative $N(\mu_0,\tau_0)$ prior for
        $\beta_0$ in the case where $\alpha=0$. This will ensure that
        the probit and the skewed-probit models have comparable priors
        for their respective "intercept" parameters.
        
        \subsection{Issue 3 - Skewness-related parameters}\label{skewness_section}

        It is well-known that the skew-normal likelihood has a
        (double) singularity in the neighbourhood $\alpha \simeq 0$
        \cite{azzalini1985}. Various adaptations of maximum likelihood
        estimation and some Bayes estimators have been proposed as
        solutions to this singularity. \cite{hallin2014} used the
        Fisher information to propose a reparameterization that uses
        $\alpha^3$ as the skewness parameter since this solves the
        double singularity problem in the likelihood. In our venture
        to derive the PC prior for the skewness, we derived the
        Kullback-Leibler divergence (KLD) from the skew-normal link to
        the probit link and noticed the same feature as mentioned in
        \cite{hallin2014}. This resemblance is expected since the
        Fisher information metric is the Hessian of the KLD.

        From \eqref{skewnormpdf2}, the KLD for small $|\alpha|$ can be
        found to be
        \begin{eqnarray}
          \text{KLD}(\alpha) &=& \int f(x|\alpha)\log
                                 \frac{f(x|\alpha)}{f(x|\alpha=0)}dx\notag\\
                             &= & \frac{\pi^2 + 16 - 8\pi}{6\pi^3}\alpha^6
                                  -\frac{144\pi + 3\pi^3-38\pi^2-168}{6\pi^4}\alpha^8\notag\\
                             && +  \frac{-42240\pi - 2560\pi^3+16176\pi^2
                                +129\pi^4+39936}{120\pi^5}\alpha^{10} +
                                \mathcal{O}(\alpha^{12})\notag \\
                             &\approx&c_1\alpha^6 + c_2\alpha^8 + c_3\alpha^{10}.
                                \label{kld2}
        \end{eqnarray}
        Interestingly, the behavior of $\alpha$ around $\alpha=0$ does
        not have the usual asymptotics (consistency rate of
        $\sqrt{n}$) since the leading term is $\alpha^6$. This implies
        that the estimator of $\alpha$ in the neighbourhood
        $\alpha\simeq 0$, has a consistency rate $n^\frac{1}{6}$ but a
        skewness parameter $\gamma = \alpha^3$, such that
        $\alpha = \text{sign}(\gamma)\sqrt[3]{|\gamma|}$, will have
        the normal asymptotics in the sense that the estimator of
        $\gamma$ will be $\sqrt{n}$ consistent.
        
        Even though $\gamma$ has the usual asymptotic behaviour, the
        estimate of it is hard to interperate since it does not relate
        easily to an interpretable property. We can instead focus on
        the more intepretable (standarised) skewness of the
        skew-normal distribution, $\gamma_1$, which is a monotone
        function of $\gamma$
        \begin{equation}
            \gamma_1 =
            \frac{(4-\pi)\left(\sqrt{\frac{2\delta^2}{\pi}}\right)^3}{%
                2(1-\frac{2\delta^2}{\pi})^\frac{3}{2}},
        \end{equation}
        where $\delta = \frac{\alpha}{\sqrt{1+\alpha^2}}$ (and
        $\gamma=\alpha^{3}$). The skewness take values in the interval
        $-0.99527<\gamma_1<0.99527$, which is correct up to five
        digits.
        
        The question arises if we should formulate a prior for
        $\alpha$, $\gamma$ or the skewness $\gamma_1$. If priors are
        assigned more ad-hoc parameters, this question poses a
        challenge. The PC prior is invariant under reparameterizations
        \cite{simpson2017}, implying that this framework will produce
        equivalent priors for $\alpha$, $\gamma$ and $\gamma_1$. They
        are equivalent in the inferential sense, and will produce the
        same posterior inference.
         
         \section{SKEW-NORMAL MEAN REGRESSION}\label{skewregsec}
         In this section we focus on skew-normal regression, although 
         these issues also exist in more general skew-symmetric regression
          models. 
          
         In the preceeding section we mentioned the different parameters
          that can be used to capture the skewness in the skewed probit 
          model, and the proposals pertain to the skew-normal regression
           model as well. 
           
           Most works on skew-normal regression propose a regression model for the location parameter, $\xi$, from \eqref{skewnormpdf}. This generalization of Gaussian regression seems straightforward but when we keep in mind that the location parameter of the Gaussian is equal to the mean, then we can see that regressing through the location parameter of the skew-normal is not practical. In the spirit of generalizing Gaussian regression to skew-normal regression, we should formulate the regression model based on the mean. Hence for $y_i\sim SN(\xi,\omega,\alpha)$ from \eqref{skewnormpdf},
           \begin{eqnarray}
           E[Y_i] =  \eta_i, 
           \end{eqnarray}
           with $\eta_i$ from \eqref{linpred}, instead of $\xi_i = \eta_i$. Note that here we do not reformulate the intercept as in Section \ref{secint} for skewed probit regression, since the identity link function is used. 
           We illustrate the proposed skew-normal regression model in Section \ref{realsection}.
                 
        \section{PENALIZING COMPLEXITY PRIOR FOR THE SKEWNESS PARAMETER}\label{pcsection}

        The work of \cite{simpson2017} introduced the notion of
        penalizing complexity priors for parameters and provided the
        framework for deriving priors that quantify the contraction
        from a complex model to a simpler model. These PC priors are
        especially helpful and very needed in cases where priors have
        traditionally been chosen due to mathematical convenience, or
        convention.
        
        In this section we derive the PC prior for $\alpha$ due to the
        invariance of the PC prior under reparameterization of the
        skewness parameter. The derivations of the PC prior for
        $\gamma$ and $\gamma_1$ follows then directly from a
        change-of-variable exercise.

        Using \cite{simpson2017} and \eqref{kld2}, define the
        uni-directional distance from the skew-normal to the Gaussian
        density as,
        \begin{eqnarray}
          d(\alpha)&=&\sqrt{2\text{KLD}(\alpha)}\notag\\
                   &\approx& \sqrt{2(c_1\alpha^6 + c_2\alpha^8 + c_3\alpha^{10})}
                             \label{distsn}
        \end{eqnarray} 
        The penalizing complexity prior for the skewness parameter
        $\alpha$ is then formed by assigning an exponential prior with
        parameter $\theta$ to the distance. The parameter $\theta$
        incorporates information from the user to control the tail
        behavior and thus the rate of contraction towards the probit
        link function. The penalizing complexity prior follows then
        directly, as
        \begin{eqnarray}
          \pi(\alpha)&=&\frac{1}{2}\theta\exp\left[-\theta
                         d(\alpha)\right]
                         \left|\frac{\partial d(\alpha)}{\partial \alpha}\right|\notag\\
                     &\approx&\frac{\theta}{2\sqrt{2(c_1\alpha^6 +
                               c_2\alpha^8 + c_3\alpha^{10})}}
                               \left|2(6c_1\alpha^5 + 8c_2\alpha^7 +
                               10c_3\alpha^{9})\right|\notag\\
                     &&\times\exp\left[-\theta|\alpha^3|\sqrt{2(c_1 +
                        c_2\alpha^2 + c_3\alpha^{4})}\right].\label{pcalpha}
        \end{eqnarray}
        for small values of $|\alpha|$. The user-defined parameter
        $\theta$ is used to govern the contraction towards probit
        regression, e.g., for small $p_U>0$,
        \begin{eqnarray*}
          \text{Prob}(d(\alpha)>U)&=&p_U=\exp(-\theta U)
        \end{eqnarray*}
        which gives $\theta=-{\log p_U}/{U}$. There is no explicit
        expression for the penalizing complexity prior of $\alpha$ in
        general, but the prior can be computed numerically. The prior
        for $\gamma_1$ is available in the \textit{R-INLA}
        package\cite{rue2009} with \texttt{prior = "pc.sn"} and
        parameter \texttt{param=$\theta$}. We use the
        $\gamma_1$ reparameterization, since
        $\gamma_1$ quantifies the skewness as a \emph{property} with
        good interpretation.

        The PC priors of $\alpha$ and $\gamma_1$ are illustrated in
        Figure \ref{pcfigtheta5} for $\theta = 5$, on the $\alpha$ and
        $\gamma_1$ scales.
        In Figure \ref{figpcmany} various values for $\theta$ are
        considered to provide an intuition about the effect of
        $\theta$.
        \begin{center}
                \centering
                \begin{figure}[h]
                        \includegraphics[width= 7cm]{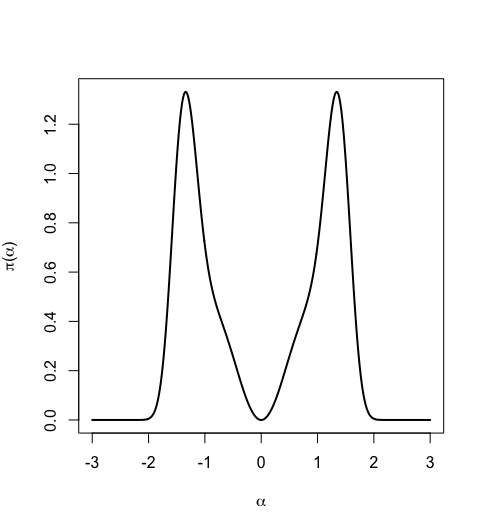}
                        \includegraphics[width= 7cm]{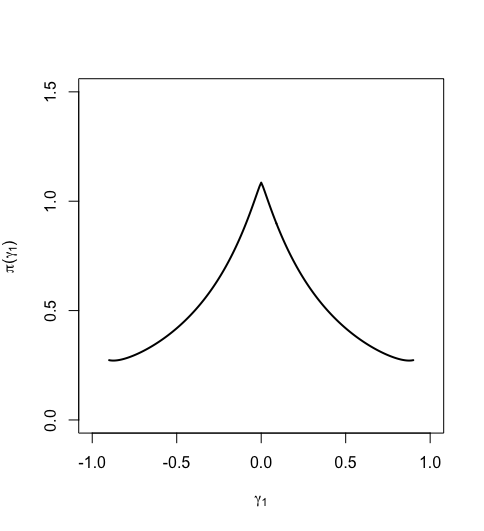}
                        \caption{PC prior \eqref{pcalpha} for
                            $\theta=5$ on the $\alpha$ scale (left)
                            and the $\gamma_1$ scale (right)}
                        \label{pcfigtheta5}
                \end{figure}
        \end{center}
        From Figure \ref{pcfigtheta5} we can see the shape of the PC
        prior for $\alpha$ is quite peculiar, but has a clear
        interpretation in terms of a prior on the distance. It just
        shows that if we assign priors to parameters, like $\alpha$,
        instead of to a property, like $\gamma_1$, it is highly
        improbable that we could think of a density function for the
        parameter that has good translatable properties. Another
        interesting note is that from the prior density of $\alpha$
        around $\alpha=0$, we can see that most priors of $\alpha$
        proposed in literature actually results in underfitting,
        instead of the usual overfitting, since they assign too much
        density to the neighborhood around $\alpha=0$. Conversely, the
        PC prior of $\gamma_1$ is as expected with a mode at the value
        for the probit link.
        
        \begin{center}
                \centering
                \begin{figure}[h]
                        \includegraphics[width= 7cm]{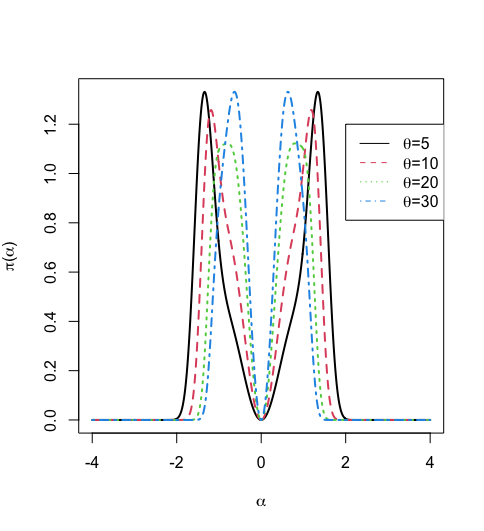}
                                \includegraphics[width= 7cm]{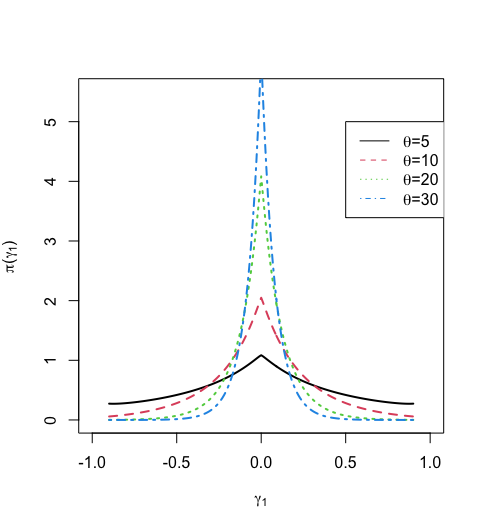}
                        \caption{PC prior \eqref{pcalpha} for various
                            $\theta$'s on the $\alpha$ scale (left)
                            and the $\gamma_1$ scale (right)}
                        \label{figpcmany}
                \end{figure}
        \end{center}

        \section{SIMULATION STUDY}\label{simstudy}

        In this section we present condensed results from a simulation
        study with the aim to show the results proposed in this work
        for experiments with a large and small number of trials. The
        setup is to simulate linear predictors
        $\eta_i = \beta_0(\alpha,q) + \beta_1x_i$, where
        $x_i\sim \text{N}(0,0.5)$ for $i = 1,\ldots, n$. The success
        probabilities are then $p_i = F(\eta_i|\alpha)$ from
        \eqref{skewprobitgen} and subsequently the response variable
        $y_i$, wherere $y_i\sim \text{Bin}(N_i,p_i)$. Throughout, we assume $\theta = 5$ for the PC prior and a weak Gaussian prior with parameters $(0,10^2)$ for the skewness.
        
        \subsection{Large number of trials}

        For an experiment that consists of a large number of trials,
        we consider four simulation scenario's which can be summarized
        as:
        \begin{enumerate}
                \item $q = \frac{1}{3}, \beta_1 = 1, \gamma_1 = 0 (\alpha=0), N_i = 200$
                \item $q = 0.25, \beta_1 = -1, \gamma_1 = \frac{2}{3} (\alpha=10), N_i = 200$
                \item $q = 0.30, \beta_1 = 1, \gamma_1 = \frac{1}{3} (\alpha=2), N_i = 200$
                \item $q = 0.10, \beta_1 = -1, \gamma_1 = -\frac{1}{3} (\alpha=-2), N_i = 200$
        \end{enumerate}
        In each case we consider the PC prior as well as the Gaussian
        prior for the skewness $\gamma_1$, and weakly informative
        Gaussian priors for the fixed effects.

        \subsubsection{Results}

\begin{center}
                \begin{figure}[h]
                        \includegraphics[width=3.3cm]{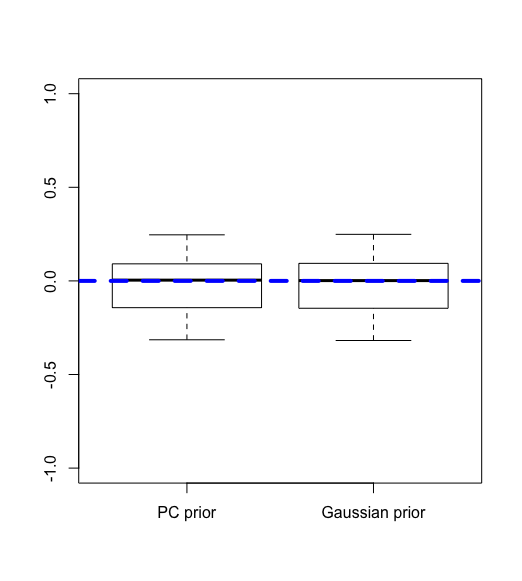}
                        \includegraphics[width=3.3cm]{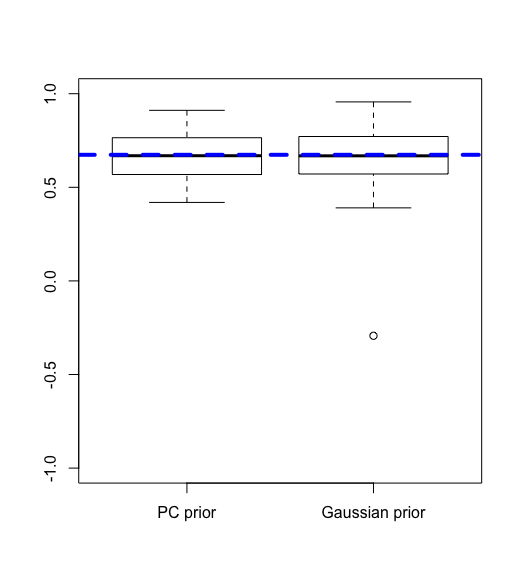}
                        \includegraphics[width=3.3cm]{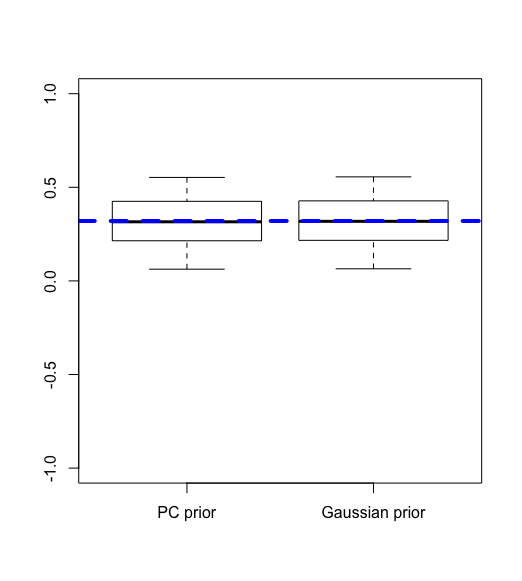}
                        \includegraphics[width=3.3cm]{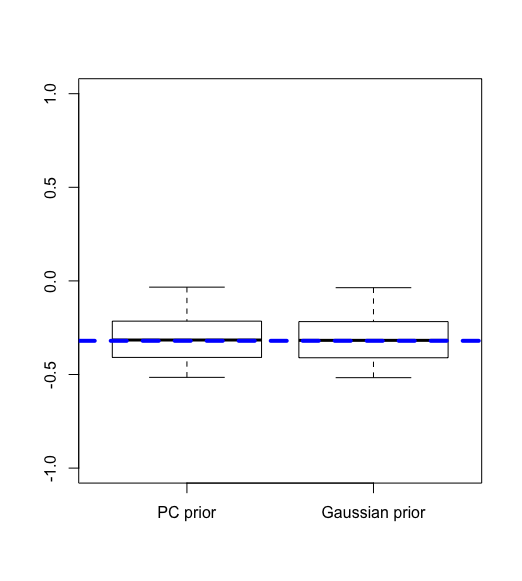}
                        \caption{Median of $95\%$ credible intervals
                            for the different scenario's with the true
                            skewness (dashed line): Scenario 1, 2, 3
                            and 4 from left to right}
                        \label{plotsim1}
                \end{figure}
        \end{center}

        The fixed effects were recovered well and here we focus on the
        skewness $\gamma_1$. From Table \ref{tabressim1} it is clear
        that the PC prior (and the Gaussian prior) performs as
        expected since the sample size and number of trials are large. 
        In Figure \ref{plotsim1} the posterior results for the
        skewness are summarised with coverage probability and median
        length of the credible interval. The results for other
        scenarios are similar and omitted here. From this (and many
        other) simulation studies, we conclude that for a large number
        of trials the skewed-probit link works well and the inference
        is accurate. It is clear that the PC prior does not contract towards the probit model when the data presents strong support for the skewed probit model (scenarios 2,3 and 4).
        \begin{table}[h]
                \begin{center}
                        \begin{tabular}{|c||c|c||c|c||}
                                \hline
                                \multirow{2}{*}{\textbf{Scenario}} 
                          & \multicolumn{2}{c||}{\textbf{PC prior}} 
                          & \multicolumn{2}{c||}{\textbf{Gaussian prior}} \\ \cline{2-5}
                          & \textbf{CP} & \textbf{MLCI}
                          & \textbf{CP} & \textbf{MLCI}\\ \hline \hline
                          \pmb{1} & $95$ & $0.28$ & $94$ & $0.35$ \\ \hline
                          \pmb{2} & $96 $ & $0.28$ & $ 97$ & $0.34$ \\ \hline
                          \pmb{3} & $95 $ & $0.31$ & $95 $ & $0.34$ \\ \hline
                          \pmb{4} & $95 $ & $0.32$ & $ 95$ & $0.35$ \\ \hline
                        \end{tabular}
                        \caption{Coverage probability (CP) and median
                            length of the credible
                            interval (MLCI) for the skewness
                            $\gamma_1$ under the PC
                            and Gaussian (G) priors, for large $N_i$}
                        \label{tabressim1}
                \end{center}
        \end{table}
        
        \subsection{Small number of trials}

        Here we focus our attention on samples of size $200$ of binary
        trials, and the scenario's we consider are:
        \begin{enumerate}
                \item $q = \frac{1}{2}, \beta_1 = 1, \gamma_1 = -\frac{2}{3} (\alpha=-10), N_i = 1$     
                \item $q = \frac{1}{2}, \beta_1 = 1, \gamma_1 = 0 (\alpha=0), N_i = 1$
        \end{enumerate}
        We consider the PC prior as well as the Gaussian prior for the
        skewness parameter, and weakly informative Gaussian priors for
        the fixed effects.

        \subsubsection{Results}

\begin{figure}
                \includegraphics[width = 4.4cm]{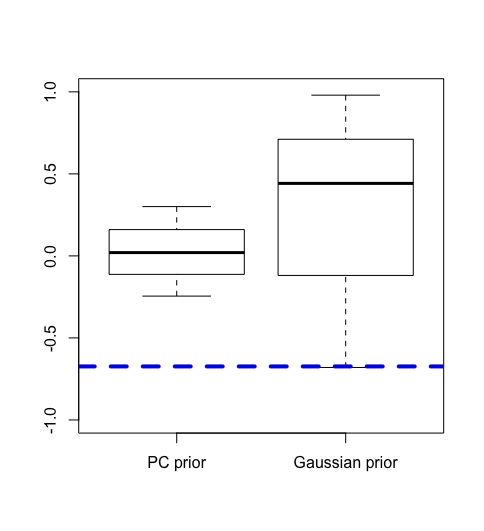}
                \includegraphics[width = 4.4cm]{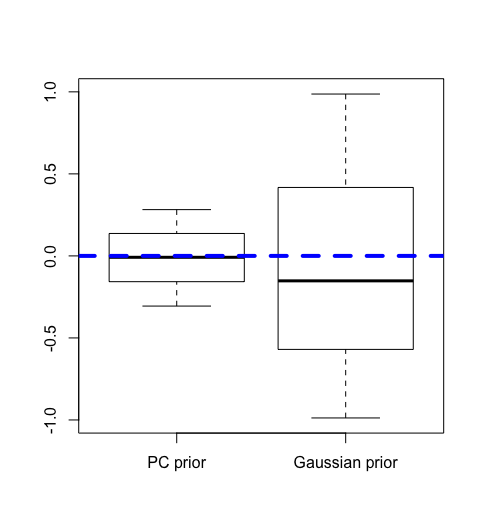}
                \includegraphics[width = 4.3cm]{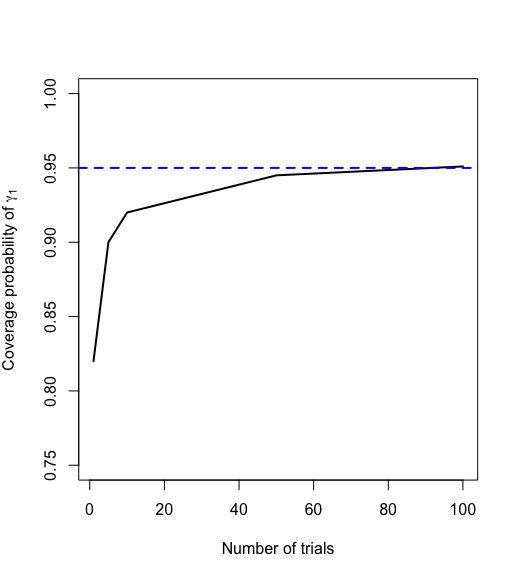}
                \caption{$95\%$ credible intervals for $\gamma_1$
                    with $n_i=1$ and $\gamma_1=-\frac{2}{3}$ (left) or
                    $\gamma_1=0$ (middle). Coverage probabilities for
                    $\gamma_1$ under scenario 1 as $n_i$ increases (right)}
                \label{trialsplot}
        \end{figure}

        From Table \ref{tabressim} it is clear that the skewness is
        not recovered well for a small number of trials. In the case
        of the PC prior, the coverage is poor but the credible
        intervals are still relatively narrow. For the Gaussian prior,
        the coverage is high mainly due to the very wide credible
        intervals. For a small number of trials or binary trials, the
        skewness is hard to capture. Even though the nominal coverage
        for the Gaussian prior is still high from Table
        \ref{tabressim}, the median length of the credible interval
        implies that the credible intervals span most of the support
        of $\gamma_1$. However, the PC prior contracts to zero with
        relatively narrow credible intervals and exhibits poor
        coverage for $\gamma_1\neq 0$. It is evident that the skewness
        is hard to estimate with a small number of trials. This is not
        unexpected since in binary data, we only observe a success or
        failure for each subject and subsequently the data does not
        provide sufficient information about the skewness. We need
        repetitions in the data to learn more about the skewness. We
        can see in Figure \ref{trialsplot} that the PC prior contracts
        to zero if there is not enough evidence for the skewed link,
        but the Gaussian prior proposes an arbitrary value for the
        skewness from most of the range of $\gamma_1$ (possibly with
        the wrong sign as in Figure \ref{trialsplot}]). In this case,
        using the skewed-probit link for binary data might not be
        useful.

        \begin{table}[h]
                \begin{center}
                        \begin{tabular}{|c||c|c||c|c||}
                                \hline
                                \multirow{2}{*}{\textbf{Scenario}} 
                          & \multicolumn{2}{c||}{\textbf{PC prior}} 
                          & \multicolumn{2}{c||}{\textbf{Gaussian prior}} \\ \cline{2-5}
                          & \textbf{CP}
                          & \textbf{MLCI}
                          & \textbf{CP}
                          & \textbf{MLCI}\\ \hline \hline
                          \pmb{1} & $65 $ & $0.41$ & $ 90$ & $1.24$ \\ \hline
                          \pmb{2} & $95 $ & $0.33$ & $ 90$ & $1.45$ \\ \hline
                        \end{tabular}
                        
                        \caption{Coverage probability (CP) and median
                            length of the credible interval (MLCI) for
                            the skewness $\gamma_1$ under the PC and
                            Gaussian (G) priors, for small $N_i$}
                        \label{tabressim}
                \end{center}
        \end{table}
        
        \subsection{Confounding and the effect of the quantile
            intercept}\label{subsecconfounding}

\begin{figure}[h]
            	\begin{center}
                \includegraphics[width=10cm]{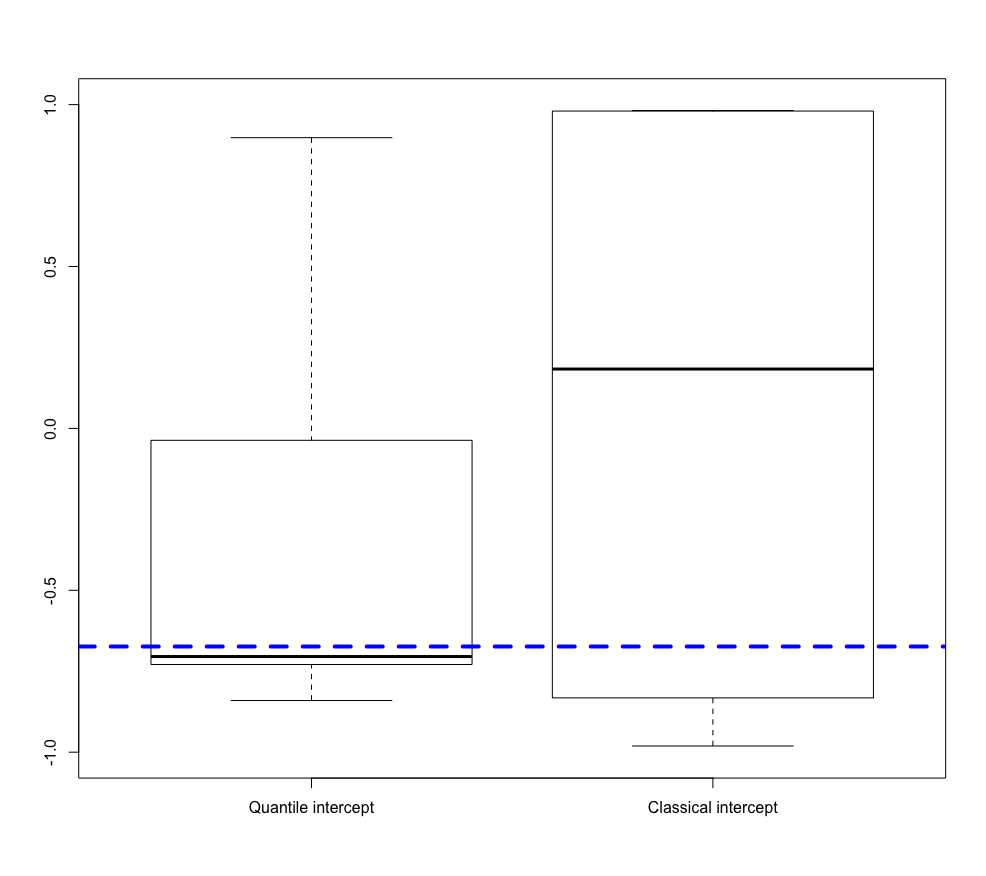}
                       \end{center}
                \caption{Median credible intervals for the skewness
                    $\gamma_1$ using the
                    quantile intercept vs the classical intercept}
                \label{difint}
            \end{figure}

        In this section we look at the effect of not using the new
        quantile intercept. We used a simulated dataset, similar to
        the preceeding section, with
        $q=0.4, \beta_1=0.1, \gamma_1=-\frac{2}{3}$. In this setup the
        linear predictor is close to zero, for a centered covariate,
        the confounding between the classical intercept and the
        skewness parameter is clear. In Figure \ref{difint} the median
        of the $95\%$ credible intervals of the skewness (for 500
        repetitions) as well as the true value of the skewness are
        presented. On the left we have the case of the quantile
        intercept and on the right, the classical intercept. By using
        the classical intercept, as in the case of GLM, the skewness
        is not estimated correctly in the sense that the direction is
        not even recovered. It is clear that the quantile intercept
        solves the confounding of the intercept of the linear
        predictor, with the skewness of the link.
        
        \section{APPLICATIONS}\label{realsection}

        In this section we illustrate the use of skewed probit
        regression with the PC prior using two well-known datasets,
        the beetle mortality data \cite{bliss1935} (binomial response
        with multiple trails) and the UCI Cleveland heart disease data
        \cite{ucidata} (Bernoulli response). We also present the analysis
         of the Wines data to illustrate the use of this work in the 
         skew-normal likelihood.

        \subsection{Beetle mortality data}

\begin{center}
                \begin{figure}[h]
                        \includegraphics[width=12cm, height = 8cm]{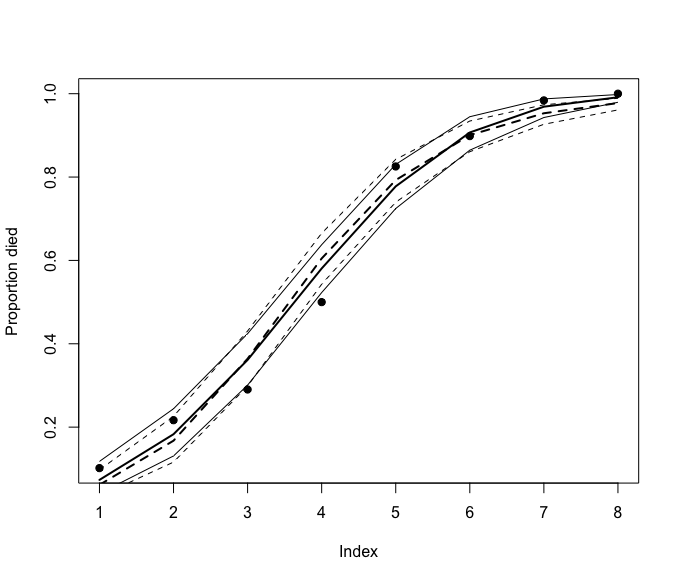}
                        \caption{Fitted and observed proportions (-- Skewed Probit, - - Probit) with $95\%$ credible intervals}
                        \label{beetlefig}
                \end{figure}
        \end{center}

        In this well-known dataset from \cite{collet2003} the number
        of adult flour beetles killed by differing dosages of poison
        is modelled based on the centered dosage value. We use the
        proposed skewed probit model with the PC prior and the
        quantile intercept. We also fit a probit model and compare the
        fitted values of both with the observed data. These, together
        with the $95\%$ credible intervals are presented in Figure
        \ref{beetlefig}. We note that the skewed probit model seem to
        fit the observed data better than the probit model, and the
        $95\%$ credible interval for the skewness of the skewed probit
        model from Table \ref{beetletabel} does not include $0$. The
        marginal log-likelihood for the skewed probit model is
        $-21.75$ versus $-23.93$ from the probit model.
The
        difference between the marginal log-likelihoods does not
        provide a convincing argument in favor of the skewed probit
        model, as opposed to the probit model.
        \begin{table}[h]
                \begin{tabular}{|l||l|l|}\hline
                        \textbf{Effect} & \textbf{Estimate} & \textbf{$\mathbf{95\%}$ credible interval}\\ \hline \hline
                        \textbf{Quantile of the intercept ($q$)} & $0.643$ & $(0.572;0.703)$ \\ \hline
                        \textbf{Dosage} & $19.132$ & $(16.074;22.316)$ \\ \hline
                        \textbf{Skewness ($\gamma_1$)} & $-0.456$ & $(-0.848;-0.053)$ \\ \hline
                \end{tabular}
                
                \caption{Posterior estimates for the beetle mortality data}
                \label{beetletabel}
        \end{table}
        
        \subsection{Heart disease data}

        We will use the Cleveland data obtained by Robert Detrano from
        the V.A. Medical Center, Long Beach and Cleveland Clinic
        Foundation.
        
        The response is a binary observation indicating the occurrence
        of a $> 50\%$ diameter narrowing in an angiography. Various
        covariates are available in this data and we will use a subset
        of these namely, gender (male/female), type of chest pain (1 -
        typical angina, 2 - atypical angina, 3 - non-anginal pain, 4 -
        asymptomatic), resting blood pressure, the slope of the peak
        exercise ST segment (1 - upsloping, 2 - flat, 3 - down
        sloping), the number of colored vessels by fluoroscopy and the
        results from the thallium heart scan (3 - normal, 6 - fixed
        defect, 7 - reversable defect). We centered the two continuous
        covariates, resting heart rate and the number of colored
        vessels by fluoroscopy. Further details can be found in
        \cite{ucidata}.
        
        There are 297 subjects with complete information in the
        dataset of which 137 experienced the event of $> 50\%$
        diameter narrowing in an angiography. We fit a skewed-probit
        regression model to explain the probability of the event based
        on the values of the covariates similar to \cite{lee2019}. In
        \cite{lee2019} divergent results were obtained based on
        different estimation frameworks, namely maximum likelihood
        estimation, bootstrap bias correction, Jeffrey's prior,
        generalized information matrix prior and Cauchy prior
        penalized frameworks. The inconsistent results could be
        attributed to the issues we mentioned in this paper, since all
        these estimation methods were developed for the skewed-probit
        regression model without the good standardization, based on
        the skewness parameter $\alpha$ and defined using the
        classical intercept.
        
        Also, there is a lack of information on the skewness in binary
        data. The consequence is thus that various values of the
        skewness could be supported. This setup necessitates the need
        for the PC prior of the skewness, so that we will use probit
        regression unless the data strongly supports skewed probit
        regression.
        
        Here, we can use the PC prior \eqref{pcalpha} for the skewness
        and the quantile intercept from Section \ref{secint}. All
        quantitative covariates are centered. The results are given in
        Table \ref{tabheart}.
        
        \begin{table}[h]
                \begin{tabular}{|l||l|l|}
                        \hline
                        & \textbf{Posterior mean} & \textbf{$\mathbf{95\%}$ credible interval}\\ \hline \hline
                        \textbf{Quantile Intercept ($q$)} & $0.045$ & $(0.006;0.184)$\\ \hline 
                        \textbf{Gender (male)} & $1.025$ & $(0.605;1.461)$\\ \hline 
                        \textbf{Type of chest pain (2)} & $0.198$ & $(-0.538;0.942)$\\ \hline 
                        \textbf{Type of chest pain (3)} & $-0.074$ & $(-0.732;0.590)$\\ \hline 
                        \textbf{Type of chest pain (4)} & $1.288$ & $(0.673;1.920)$\\ \hline 
                        \textbf{Resting heart rate} & $0.016$ & $(0.005;0.027)$\\ \hline 
                        \textbf{Slope of the peak exercise (2)} & $1.027$ & $(0.637;1.452)$\\ \hline 
                        \textbf{Slope of the peak exercise (3)} & $0.791$ & $(0.059;1.540)$\\ \hline
                        \textbf{Number of colored vessels} & $0.704$ & $(0.477;0.945)$\\ \hline  
                        \textbf{Skewness $(\gamma_1)$} & $0.02$ & $(-0.214;0.235)$\\ \hline
                \end{tabular}
                \caption{Results for the Cleveland heart disease data}
                \label{tabheart}
        \end{table}

\begin{figure}[h]
                \includegraphics[width=6cm]{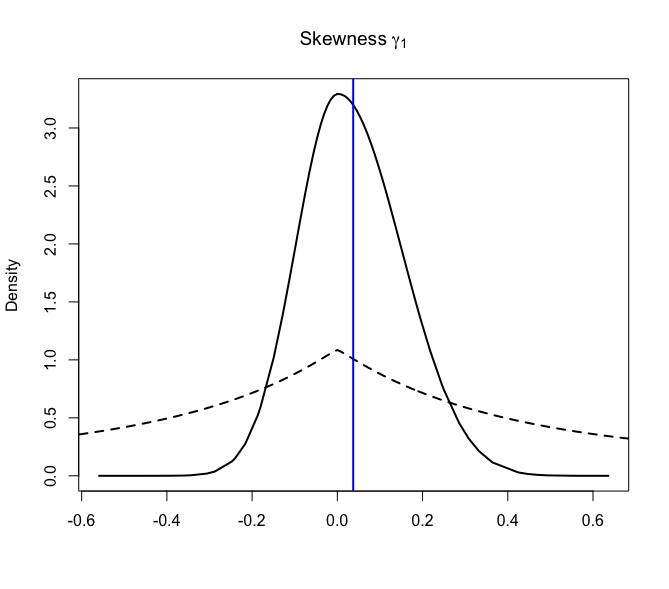}
                 \includegraphics[width=6cm]{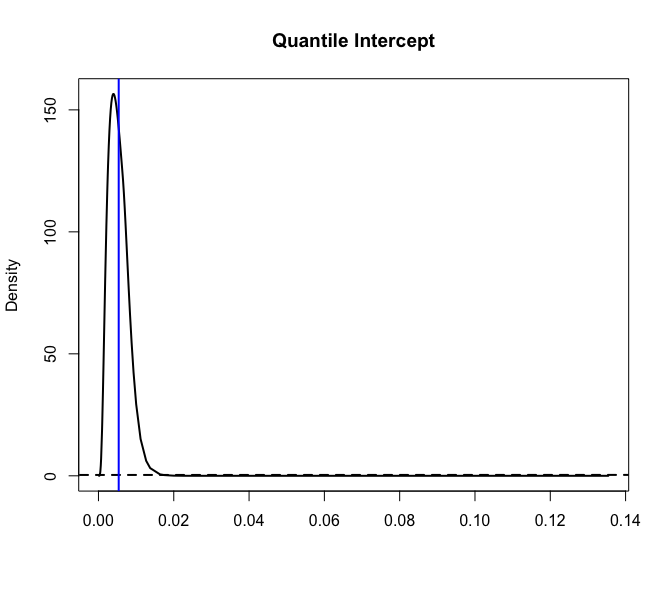}
                \caption{Posterior (prior - dashed) density of the
                    skewness $\gamma_1$ (left) and quantile intercept $q$ (right) with the corresponding point
                    estimates (vertical line)}
                \label{figskewheart}
        \end{figure}

        From the estimate of $\gamma_1$ in Table \ref{tabheart} we
        deduce that the skewness is not supported by the data and a
        probit regression model could be sufficient. We did the
        analysis using probit regression and the inference is very
        similar. This result coincides with most of the results in
        \cite{lee2019}. The posterior densities (and prior densities in
        dashed) of the skewness, $\gamma_1$, and quantile intercept, $q$, are presented in Figure
        \ref{figskewheart}.
        
        We also see that being a male, having asymptomatic chest pain,
        higher resting heart rate, a flat or downwards slope of the
        peak exercise ST segment and more colored vessels by
        fluoroscopy, all contribute to a higher probability of the
        event under investigation, i.e. $> 50\%$ diameter narrowing in
        an angiography.
        
\begin{figure}[h]
            \begin{center}
               
                \includegraphics[width=5.5cm]{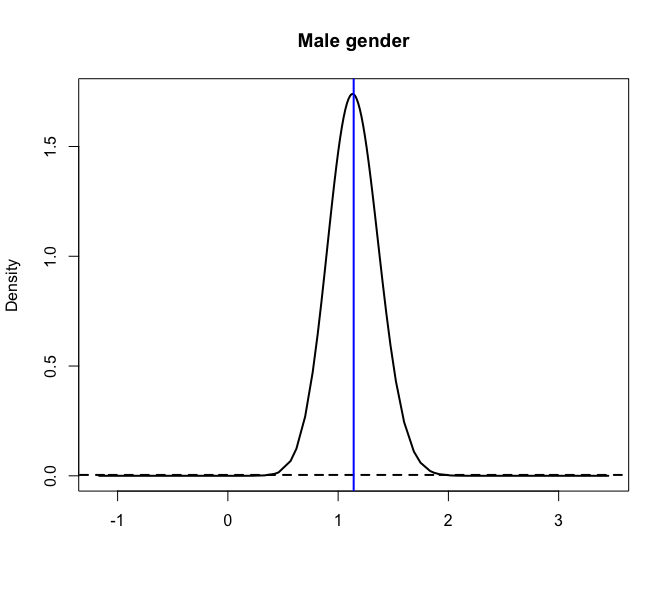}
                \includegraphics[width=5.5cm]{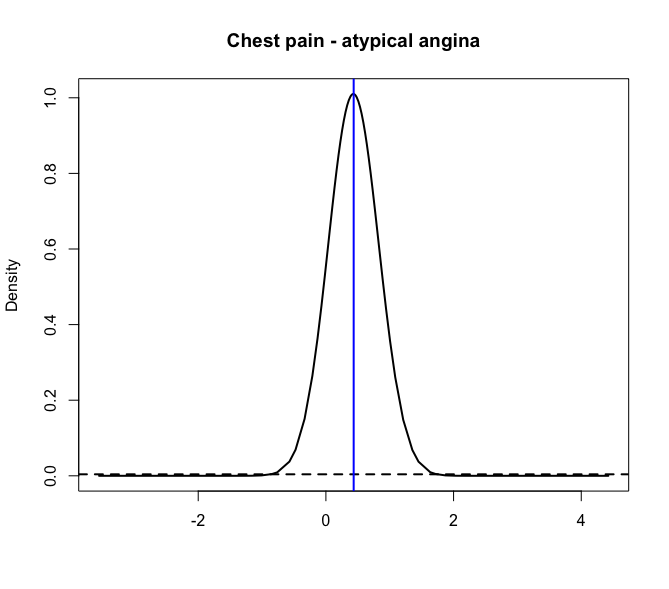}\\
                \includegraphics[width=5.5cm]{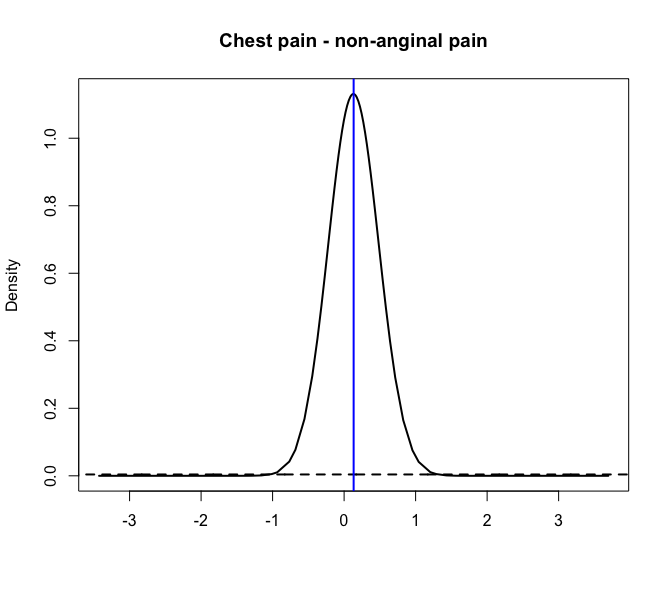}
                \includegraphics[width=5.5cm]{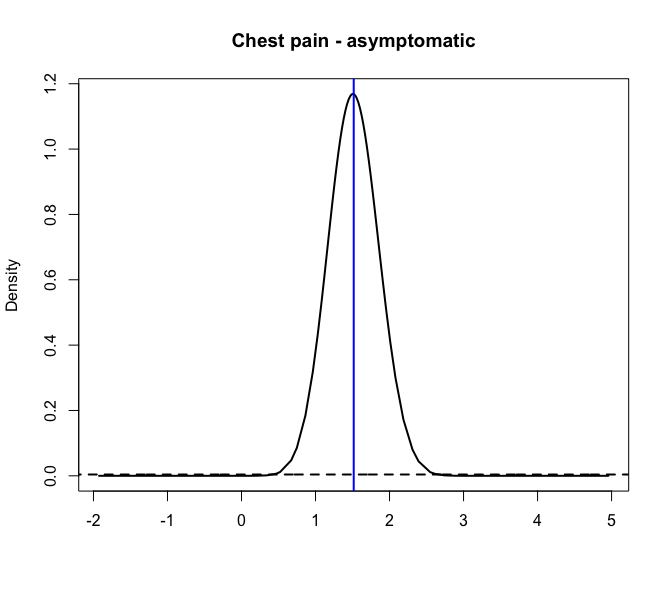}\\
                \includegraphics[width=5.5cm]{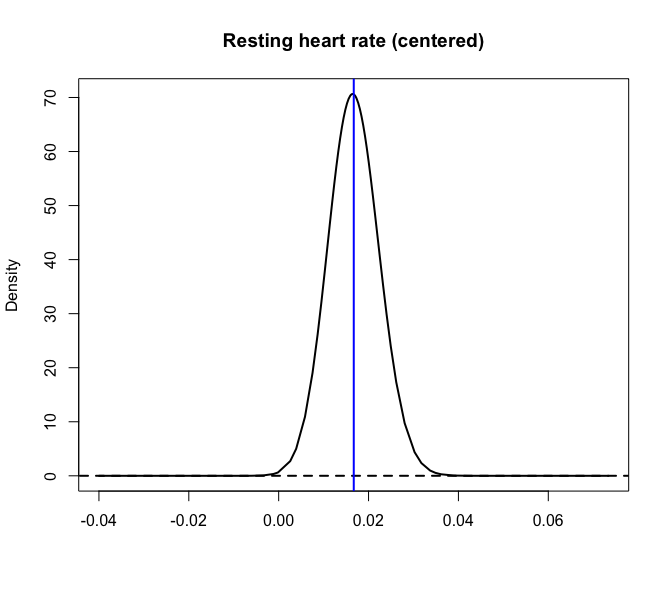}
                \includegraphics[width=5.5cm]{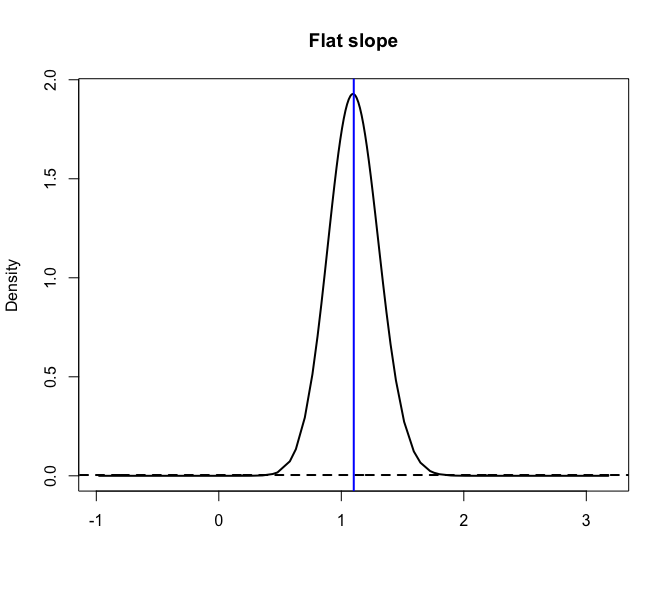}\\
                \includegraphics[width=5.5cm]{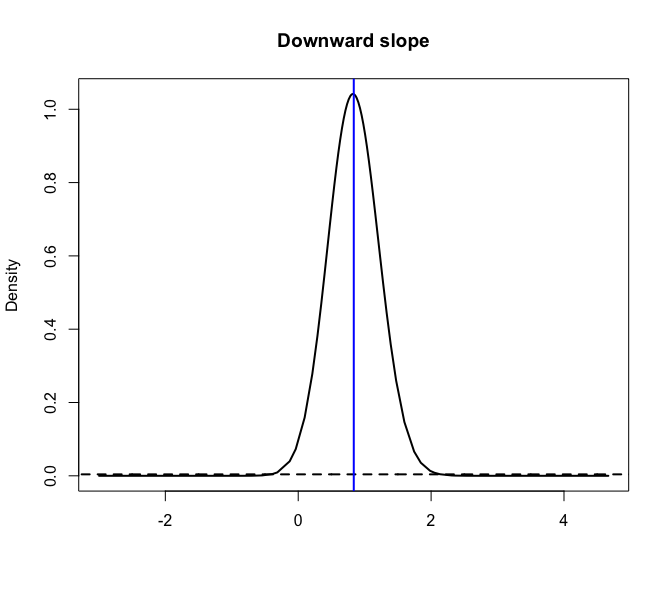}
                \includegraphics[width=5.5cm]{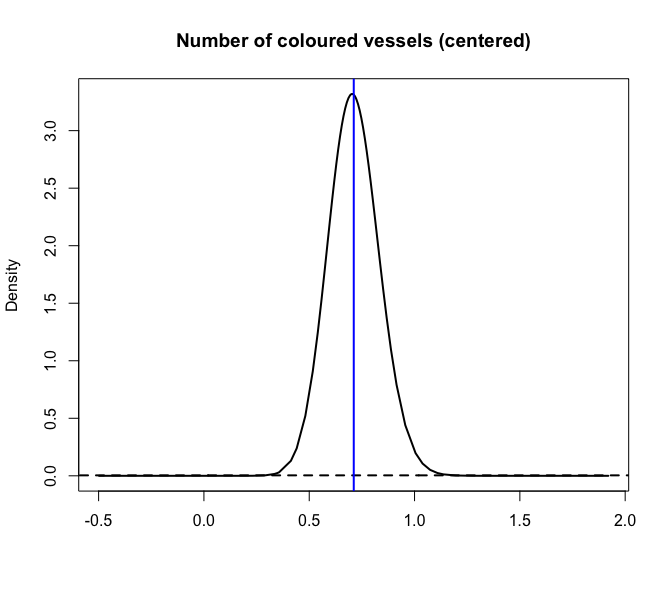}
            \end{center}
            \caption{Posterior (prior - dashed) densities of the
                    fixed effects
                    with the corresponding point estimate (vertical line)}
                \label{figpostsheart}   
        \end{figure}

        The posterior densities (and prior densities in dashed) of the
        fixed effects are presented in Figure \ref{figpostsheart}.
        
        We calculated the marginal log-likelihoods for the probit and
        skewed-probit models to be $-150.62$ and $-158.41$,
        respectively, indicating that the probit model is preferred by
        the data. Both models achieved a correct classification
        percentage of $84.55\%$, on a $50\%$ holdout sample.

		\subsection{Wines data}
\begin{center}
			\begin{figure}
				\includegraphics[width=10cm]{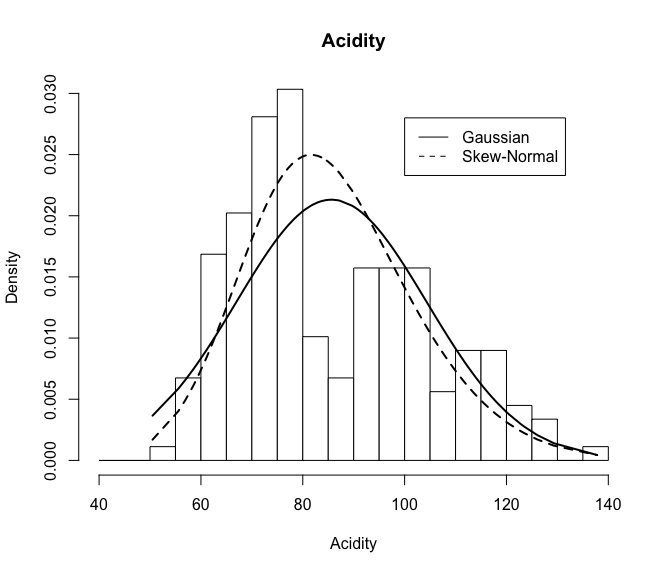}
				\caption{Histogram with model-based Gaussian curve and skew-normal curve}
				\label{figwinehist}
			\end{figure}
		\end{center}

		This section illustrates the new results when the response variable is continuous and assumed to follow a skew-normal distribution. As mentioned in Section \ref{skewregsec}, the results derived in this paper hold for skewed-probit models, as well as skew-normal regression models. We use the wines dataset from \cite{azzalini2013book}, where the acidity of the wine is assumed to follow a skew-normal distribution as illustrated in Figure \ref{figwinehist}, where we see the tail behaviour is correctly captured by the fitted Gaussian density, but not the skewness. The mean acidity (not the location parameter) is modelled using the type of wine, sugar content and pH level as covariates (after backwards elimination). We assign PC priors for the precision \cite{simpson2017} as well as skewness \eqref{pcalpha}. The results are given in Table \ref{tabwines}. The marginal log-likelihood for the skew-normal model is $-722.21$ and for the Gaussian model it is $-724.59$.

		\begin{table}[h]
			\begin{tabular}{|l||l|l|}
				\hline
				& \textbf{Posterior mean} & \textbf{$\mathbf{95\%}$ credible interval}\\ \hline \hline
				\textbf{Intercept} & $77.053$ & $(73.824;80.252)$\\ \hline 
				\textbf{Wine (Grignolino)} & $5.088$ & $(0.478;9.693)$\\ \hline 
				\textbf{Wine (Barbera)} & $23.613$ & $(19.003;28.280)$\\ \hline 
				\textbf{Sugar} & $3.118$ & $(1.150;5.080)$\\ \hline 
				\textbf{pH} & $-8.350$ & $(-10.122;-6.574)$\\ \hline 
				\textbf{Skewness $(\gamma_1)$} & $0.439$ & $(0.128;0.702)$\\ \hline
				\textbf{Precision for the data}  & $0.008$ & $(0.006;0.009)$ \\ \hline
			\end{tabular}
			\caption{Results for the wines data}
			\label{tabwines}
		\end{table}

        \section{DISCUSSION}\label{disc}

        The use of skew-symmetric distributions or links is popular
        due to the perceived flexibility inherited through the extra
        parameter that controls the skewness. The skew normal skewness
        parameter in particular, poses various challenges in the
        inference thereof. As we set out with the initial aim to
        derive the penalizing complexity prior for the skewness, we
        realized that there are various other issues that we could not
        found addressed in the literature. It is apparent that with
        the generalizing to skew-symmetric distributions and links
        from the symmetric counterparts, various fundamental concepts
        have gone amiss. 
        
        Here we rectify the formulation of the intercept in the linear
        predictor of all skew-symmetric links,
        firstly to ensure that it behaves as an intercept and secondly
        due to the confounding with the skewness parameter and fixed
        effects. We also show that the popular method of standardizing
        the skewed link function by inheriting the parameter values of
        the symmetric link, fundamentally changes the way the link
        function maps the data to the linear predictor, and we provide
        an anchored standardization approach. We believe that many of
        the contradicting works in this area can be attributed to the
        inappropriate use of the classical intercept and
        parameter-based standardization, instead of property-based
        standardization. In skew-symmetric regression models, we formulate the regression model based on the mean, instead of the location parameter.
        
        After the fundamental corrections to the formulation of the
        skewed-probit link, the penalizing complexity prior for the
        skewness was derived. One particular advantage of this prior
        is that it is invariant to reparameterizations of the skewness
        parameter. In light of this, we implemented the PC prior for
        the skewness in \textit{R-INLA} \cite{rue2009} for use by others. We
        noted, expectedly, that binary data (or with few trials) does
        not provide information about the skewness, and we thus advise
        against the use of the skewed-probit link for data with a
        small number of trials.  We advocate the use of the PC prior
        even more feverently because of this feature, since the PC
        prior will contract to the simpler probit link instead of
        providing an incorrect unreliable estimate of the skewness.
        Other inferential frameworks might not be able to ensure this
        contraction in the absence of convincing evidence from the
        data about the necessary skewness, and could lead to unfounded
        complicated models.
        
        We hope that the issues raised and addressed here will improve
        the inference of the skewed probit model (and more broadly the
        skew-symmetric links and likelihoods) and provide insights into the
        fundamental considerations necessary when distributions or
        links are generalized.
        
        \section*{Appendix}

        We give here a small example for how to do skew probit
        regression in \textit{R-INLA}. In the code below, the unusual
        statement is \texttt{remove.names="(Intercept)"} which
        remove the intercept in the formula \emph{after} doing the
        expansion of factors in the model. We need this as we replace
        the traditional intercept with the quantile intercept in the
        link, and the expansion of factors depends on the presence or
        not, of an intercept in the model.
        \begin{verbatim}
library(INLA)
n = 200
Ntrials = 200
x = rnorm(n, sd = 0.5)
eta = x
skew <- 0.5
prob = inla.link.invsn(eta, skew = skew, intercept = 0.75)
y = rbinom(n, size = Ntrials,  prob = prob)
r = inla(y ~ 1 + x,
    family = "binomial",
    data = data.frame(y, x),
    Ntrials = Ntrials,
    control.fixed = list(remove.names = "(Intercept)",
                         prec = 1), 
    control.family = list(
        control.link = list(model = "sn", 
                            hyper = list(
                                skew = list(param = 10)))))
summary(r)
\end{verbatim}

\bibliographystyle{apalike}
\bibliography{BioJ.bib}

\begin{thebibliography}{}

\bibitem[Arellano-Valle et~al., 2007]{arellano2007}
Arellano-Valle, R., Bolfarine, H., and Lachos, V. (2007).
\newblock Bayesian inference for skew-normal linear mixed models.
\newblock {\em Journal of Applied Statistics}, 34(6):663--682.

\bibitem[Azevedo et~al., 2011]{azevedo2011}
Azevedo, C.~L., Bolfarine, H., and Andrade, D.~F. (2011).
\newblock Bayesian inference for a skew-normal irt model under the centred
  parameterization.
\newblock {\em Computational Statistics \& Data Analysis}, 55(1):353--365.

\bibitem[Azzalini, 1985]{azzalini1985}
Azzalini, A. (1985).
\newblock A class of distributions which includes the normal ones.
\newblock {\em Scandinavian journal of statistics}, pages 171--178.

\bibitem[Azzalini, 2013]{azzalini2013book}
Azzalini, A. (2013).
\newblock {\em The skew-normal and related families}, volume~3.
\newblock Cambridge University Press.

\bibitem[Azzalini and Arellano-Valle, 2013]{azzalini2013}
Azzalini, A. and Arellano-Valle, R.~B. (2013).
\newblock Maximum penalized likelihood estimation for skew-normal and skew-t
  distributions.
\newblock {\em Journal of Statistical Planning and Inference}, 143(2):419--433.

\bibitem[Azzalini and Capitanio, 2003]{azzalini2003}
Azzalini, A. and Capitanio, A. (2003).
\newblock Distributions generated by perturbation of symmetry with emphasis on
  a multivariate skew t-distribution.
\newblock {\em Journal of the Royal Statistical Society: Series B (Statistical
  Methodology)}, 65(2):367--389.

\bibitem[Bayes and Branco, 2007]{bayes2007}
Bayes, C.~L. and Branco, M.~D. (2007).
\newblock Bayesian inference for the skewness parameter of the scalar
  skew-normal distribution.
\newblock {\em Brazilian Journal of Probability and Statistics}, pages
  141--163.

\bibitem[Baz{\'a}n et~al., 2010]{bazan2010}
Baz{\'a}n, J.~L., Bolfarine, H., and Branco, M.~D. (2010).
\newblock A framework for skew-probit links in binary regression.
\newblock {\em Communications in Statistics-Theory and Methods},
  39(4):678--697.

\bibitem[Baz{\'a}n et~al., 2006]{bazan2006}
Baz{\'a}n, J.~L., Branco, M.~D., Bolfarine, H., et~al. (2006).
\newblock A skew item response model.
\newblock {\em Bayesian analysis}, 1(4):861--892.

\bibitem[Bliss, 1935]{bliss1935}
Bliss, C.~I. (1935).
\newblock The calculation of the dosage-mortality curve.
\newblock {\em Annals of Applied Biology}, 22(1):134--167.

\bibitem[Cabras et~al., 2012]{cabras2012}
Cabras, S., Racugno, W., Castellanos, M.~E., and Ventura, L. (2012).
\newblock A matching prior for the shape parameter of the skew-normal
  distribution.
\newblock {\em Scandinavian Journal of Statistics}, 39(2):236--247.

\bibitem[Canale et~al., 2016]{canale2016}
Canale, A., Kenne~Pagui, E.~C., and Scarpa, B. (2016).
\newblock Bayesian modeling of university first-year students' grades after
  placement test.
\newblock {\em Journal of Applied Statistics}, 43(16):3015--3029.

\bibitem[Castro et~al., 2013]{castro2013}
Castro, L.~M., Mart{\'\i}n, E.~S., and Arellano-Valle, R.~B. (2013).
\newblock A note on the parameterization of multivariate skewed-normal
  distributions.
\newblock {\em Brazilian Journal of Probability and Statistics}, pages
  110--115.

\bibitem[Chen et~al., 1999]{chen1999}
Chen, M.-H., Dey, D.~K., and Shao, Q.-M. (1999).
\newblock A new skewed link model for dichotomous quantal response data.
\newblock {\em Journal of the American Statistical Association},
  94(448):1172--1186.

\bibitem[Collet, 2003]{collet2003}
Collet, D. (2003).
\newblock {\em Modelling Binary Data. , 2nd ed}.
\newblock Chapman \& Hall/CRC, Boca Raton , FL.

\bibitem[Czado and Santner, 1992]{czado1992}
Czado, C. and Santner, T.~J. (1992).
\newblock The effect of link misspecification on binary regression inference.
\newblock {\em Journal of statistical planning and inference}, 33(2):213--231.

\bibitem[Dette et~al., 2018]{dette2018}
Dette, H., Ley, C., and Rubio, F. (2018).
\newblock Natural (non-) informative priors for skew-symmetric distributions.
\newblock {\em Scandinavian Journal of Statistics}, 45(2):405--420.

\bibitem[Dua and Graff, 2017]{ucidata}
Dua, D. and Graff, C. (2017).
\newblock {UCI} machine learning repository.

\bibitem[Firth, 1993]{firth1993}
Firth, D. (1993).
\newblock Bias reduction of maximum likelihood estimates.
\newblock {\em Biometrika}, 80(1):27--38.

\bibitem[Genton, 2004]{genton2004}
Genton, M.~G. (2004).
\newblock {\em Skew-elliptical distributions and their applications: a journey
  beyond normality}.
\newblock CRC Press.

\bibitem[Genton and Zhang, 2012]{genton2012}
Genton, M.~G. and Zhang, H. (2012).
\newblock Identifiability problems in some non-gaussian spatial random fields.
\newblock {\em Chilean Journal of Statistics}, 3(2):171--179.

\bibitem[Hallin et~al., 2014]{hallin2014}
Hallin, M., Ley, C., et~al. (2014).
\newblock Skew-symmetric distributions and fisher information: the double sin
  of the skew-normal.
\newblock {\em Bernoulli}, 20(3):1432--1453.

\bibitem[Lee and Sinha, 2019]{lee2019}
Lee, D. and Sinha, S. (2019).
\newblock Identifiability and bias reduction in the skew-probit model for a
  binary response.
\newblock {\em Journal of Statistical Computation and Simulation},
  89(9):1621--1648.

\bibitem[Liseo, 1990]{liseo1990}
Liseo, B. (1990).
\newblock The skew-normal class of densities: inferential aspects from a
  bayesian viewpoint.
\newblock {\em Statistica}, 50:59--70.

\bibitem[Liseo and Loperfido, 2006]{liseo2006}
Liseo, B. and Loperfido, N. (2006).
\newblock A note on reference priors for the scalar skew-normal distribution.
\newblock {\em Journal of Statistical Planning and Inference}, 136(2):373--389.

\bibitem[Maghami et~al., 2020]{maghami2020}
Maghami, M.~M., Bahrami, M., and Sajadi, F.~A. (2020).
\newblock On bias reduction estimators of skew-normal and skew-t distributions.
\newblock {\em Journal of Applied Statistics}, pages 1--23.

\bibitem[O'hagan and Leonard, 1976]{ohagan1976}
O'hagan, A. and Leonard, T. (1976).
\newblock Bayes estimation subject to uncertainty about parameter constraints.
\newblock {\em Biometrika}, 63(1):201--203.

\bibitem[Otiniano et~al., 2015]{otiniano2015}
Otiniano, C., Rathie, P., and Ozelim, L. (2015).
\newblock On the identifiability of finite mixture of skew-normal and skew-t
  distributions.
\newblock {\em Statistics \& Probability Letters}, 106:103--108.

\bibitem[Rue et~al., 2009]{rue2009}
Rue, H., Martino, S., and Chopin, N. (2009).
\newblock Approximate {Bayesian} inference for latent {Gaussian} models by
  using integrated nested laplace approximations.
\newblock {\em Journal of the Royal Statistical Society: Series B (Statistical
  Methodology)}, 71(2):319--392.

\bibitem[Sartori, 2006]{sartori2006}
Sartori, N. (2006).
\newblock Bias prevention of maximum likelihood estimates for scalar skew
  normal and skew t distributions.
\newblock {\em Journal of Statistical Planning and Inference},
  136(12):4259--4275.

\bibitem[Simpson et~al., 2017]{simpson2017}
Simpson, D., Rue, H., Riebler, A., Martins, T.~G., S{\o}rbye, S.~H., et~al.
  (2017).
\newblock Penalising model component complexity: {A} principled, practical
  approach to constructing priors.
\newblock {\em Statistical Science}, 32(1):1--28.

\end{thebibliography}
        
\end{document}